%% file: paper.tex
\documentclass{adobe_research}
\usepackage[T1]{fontenc}
\usepackage{lmodern}
\usepackage{multirow}
\usepackage{soul}
\usepackage{pifont}
\usepackage[T1]{fontenc}
\usepackage[table]{xcolor}
\usepackage{graphicx}
\usepackage{amsmath,amssymb}
\usepackage{bbm}
\usepackage{makecell}
\usepackage[dvipsnames]{xcolor}
\usepackage{color, colortbl}
\usepackage{caption}
\usepackage{algorithm}
\usepackage{algpseudocode}
\usepackage{booktabs}
\usepackage{tabularx}
\usepackage{array}
\usepackage{ragged2e} 

\input{preamble}

\usepackage{etoolbox}
\makeatletter

\patchcmd{\mymaketitle}{\tcbset{top=0.5cm}}{\tcbset{top=0.22cm}}{}{}
\patchcmd{\mymaketitle}{\tcbset{bottom=0.5cm}}{\tcbset{bottom=0.22cm}}{}{}
\patchcmd{\mymaketitle}{\vskip 0.5cm}{\vskip 0.12cm}{}{}
\makeatother

\usepackage{newtxtext}

\usepackage{algpseudocode}
\definecolor{catgray}{gray}{0.92}

\usepackage[dvipsnames]{xcolor} 
\definecolor{FutureOrange}{HTML}{EC866D}

\title{StreamStudio: Exploring Video Authoring with Streaming Autoregressive Generation}

\author[1,2,\ddagger]{Boyu Li}
\author[1]{Zhe Lin}
\author[1]{Eli Shechtman}
\author[1]{Ding Li}
\author[2]{Hongbo Fu}
\author[1]{Yuqian Zhou}
\affiliation[1]{Adobe Research}
\affiliation[2]{HKUST}

\contribution[\ddagger]{Work done during an internship at Adobe Research}

\abstract{
Recent video generation models can produce increasingly complex content from simple inputs, but the way people interact with them has changed much less. In most video authoring tools, users submit a prompt, wait for a full clip, and only then inspect the result, with little opportunity to intervene and redirect the generation while it is still unfolding. We present \textbf{StreamStudio}, an interactive video authoring system built around streaming autoregressive generation. StreamStudio brings the real-time generation process directly onto a timeline, allowing users to watch a video emerge, steer its continuation, retake from an earlier moment, and edit the stream as it develops, while a canvas organizes source assets and alternative video branches. To support this, we carefully manage the inference states of autoregressive video models, including KV caches and other causal context, to preserve generation contextual continuity. StreamStudio explores a video creation workflow closer to recording: creators can follow the content as it unfolds and retake from any moment.
}

\begin{document}

\maketitle

\input{sections/1_introduction}
\input{sections/2_system_design}

\input{sections/4_implementation}

\section{Conclusion}
We present StreamStudio, a prototype system that explores a new authoring paradigm for streaming video generation. Instead of entering a prompt and waiting for a complete video before making changes, creators can interact with the generation as it unfolds: steering what comes next, returning to earlier moments, and exploring different continuations. In this way, StreamStudio shifts creative control from post-generation revision to continuous interaction with the generative process.
Our contribution centers on this authoring paradigm and the interaction mechanisms that make streaming generation directly manipulable. StreamStudio is designed as a plug-and-play interface that can work with compatible generation and editing models, allowing its interaction model to evolve alongside advances in underlying video models. As streaming models become faster and easier to control in the future, such an interface may support more flexible and controllable video creation, including richer edits and multimodal interventions during generation.

\clearpage
\newpage
\bibliographystyle{ieeetr}
\bibliography{paper}

\appendix

\end{document}

%% file: preamble.tex
\usepackage{xspace}
\usepackage{booktabs}
\makeatletter
\DeclareRobustCommand\onedot{\futurelet\@let@token\@onedot}
\def\@onedot{\ifx\@let@token.\else.\null\fi\xspace}

\makeatother

%% file: sections/1_introduction.tex
\section{Introduction}
\label{sec:introduction}

Recent advances in generative video models are reshaping video creation, allowing users to produce complex visual content from simple input (e.g., text, image, video)~\cite{dit,polyak2024movie,wan2025,hacohen2026ltx,longlive,yuan2026helios,li2025videocraft,zhou2026alive}. In response, a growing body of research and commercial systems has explored more user-friendly interfaces for generative video creation. For example, canvas-based systems integrate prompts, references, generated clips, and other media into a unified workspace~\cite{guo2026protosampling,tapnow,fireflyboards,li2026crafttrace}. However, despite increasingly sophisticated interfaces, the underlying generation workflow remains largely unchanged. Most existing video authoring tools are built around video diffusion models with bidirectional temporal attention, where users specify their intent, submit a generation request, wait for an entire clip to be synthesized, and only then inspect the result.

Such a workflow gives creators little control over a video as it is being generated. Consider a clip whose opening matches the creator's intention but whose later action does not. The creator may want to preserve the opening and change only what the character does next. He may also refine intent after seeing the generated result. Design research describes this reciprocal process, in which the problem and its solution evolve together, as the co-evolution of problems and solutions~\cite{dorst2001coevolution}. However, when feedback becomes available only after the entire video clip has been generated, creators cannot act on these evolving intentions as they emerge. Each revision requires regenerating the complete video, including content that was already satisfactory. This not only wastes computation but may also alter previously accepted portions and introduce new problems. Even with faster generation, the wait can be shortened, but it still does not change when creators can intervene or the granularity at which they can redirect the result.


\textit{What if users could instead watch a video emerge as it is being generated, and intervene immediately when they want to change its direction?} Recent progress in autoregressive video generation makes such an interaction increasingly possible. Rather than synthesizing an entire video jointly, causal video models generate frames or short chunks sequentially, allowing video to be produced as a continuous stream. Early work such as Diffusion Forcing~\cite{chen2024diffusionforcing} explored flexible sequential generation by assigning different noise levels across time, while approaches such as Self Forcing~\cite{huang2026self,cui2025self} enable real-time autoregressive video diffusion through causal rollout and KV caching. LongLive further supports real-time long-video generation guided by sequential prompts, allowing users to change prompts as generation proceeds \cite{longlive}, while editing models~\cite{shin2026motionstream,xu2026referstream} such as EditStream~\cite{zhou2026editstream} extend autoregressive streaming to a broader range of video generation and editing tasks. Together, these advances suggest a shift from generating video as a completed artifact toward generating video as an ongoing stream that can be observed and redirected. Yet, how to design an interactive creation interface that fully takes advantage of this new generation process remains largely unexplored.

We introduce \textbf{StreamStudio}, an interactive video creation system that turns streaming autoregressive generation into a human-AI co-creation workflow. StreamStudio combines a timeline for interacting with an ongoing generation process and a canvas for organizing the resulting assets and alternatives. As new frames arrive, the generated clip grows directly on the timeline, allowing users to watch the video take shape and continuously steer what comes next through updated prompts. More importantly, when the generation begins to diverge from their intent at a particular moment, users can pause the process, scrub back to that point, and continue generation from there with a new direction. Users can also select a frame and invoke streaming video-to-video editing to explore alternative visual outcomes. Each intervention creates a new video trajectory, and these trajectories are organized as connected branches on the canvas for comparison, reuse, and further exploration. StreamStudio supports text, first-frame, and reference-image conditioning through a plug-and-play architecture designed to accommodate different causal autoregressive video models.
Supporting this interaction requires preserving not only the visible frames, but also the latent inference states that maintain motion continuity and generation context. To address this challenge, we design a Git-like inference state management mechanism that checkpoints intermediate model states, including KV caches and other causal context, and restores or forks them when users retake or branch from an earlier moment. Together, we aim to make video creation feel like recording an audio track, where creators can follow the content as it unfolds and retake from any moment.

This report makes three contributions:
\begin{itemize}
    \item A video authoring paradigm in which creators work with a developing stream, guide its continuation through prompts, and revisit recoverable points to explore alternative branches.
    \item StreamStudio, a research prototype that combines a canvas with a growing timeline for viewing generated content, managing assets and branches, and performing retakes and video edits.
    \item An inference-state management mechanism for historical retakes, using bounded snapshots, reusable storage, and shared prefix command records to restore and replay generation history while preserving existing footage and inference context.
\end{itemize}

%% file: sections/2_system_design.tex
\section{System Design}

\subsection{Example Workflow}
\label{sec:example-workflow}

\begin{figure}
    \centering
    \includegraphics[width=1\linewidth]{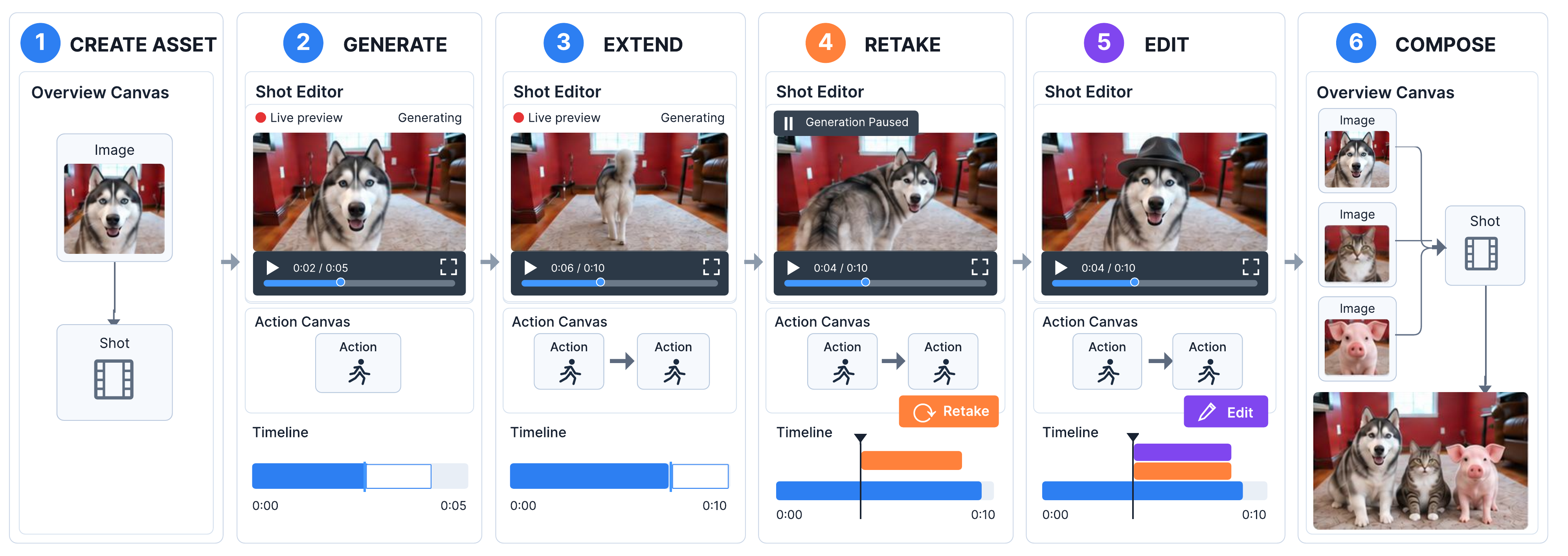}
    \caption{The step-by-step workflow of our example.}
    \label{fig:example}
\end{figure}

To illustrate the StreamStudio workflow, consider Eric, who wants to create a short video featuring a Siberian husky (Fig.~\ref{fig:example}). He begins in the \textit{Overview Canvas}, where he can create and connect visual assets and video shots.

\paragraph{Creating an image and a shot in the overview canvas.}
Eric double-clicks an empty area of the canvas and creates an \textit{Image} node. He enters the prompt ``\textit{a Siberian husky with gray-and-white fur}'' and generates a reference image of the dog. He then double-clicks the canvas again to create an empty \textit{Shot} node and connects the Image node to it. This connection assigns the generated dog image as the initial visual condition for the shot.
Eric double-clicks the Shot node to open the \textit{Shot Editor}. The project-level canvas is replaced by a focused video-authoring workspace containing a video preview, a multi-track timeline, and a smaller \textit{Shot Canvas} in the upper-left corner. While the Overview Canvas organizes assets and shots across the project, the Shot describes how the selected shot develops over time.

\paragraph{Enter shot design interface and generate.}
The Shot Canvas initially contains an \textit{Action} node that displays the connected dog image as its first-frame condition. Eric describes the opening action with the prompt:
``\textit{The Siberian husky stands in a quiet living room, blinking and gently moving its head.}''
After Eric clicks \textit{Generate}, the Action node enters a generating state. A clip appears on the timeline and gradually extends as new frames arrive. At the same time, the preview shows the husky blinking and making subtle idle movements. Eric can inspect the motion while the video is being produced rather than waiting for the entire clip to finish.

\paragraph{Continuing with a new action.}
When Eric is ready to move the scene forward, he hovers over the first Action node and clicks the displayed plus button. StreamStudio creates a connected Action node for the next part of the shot. Eric enters:
\textit{
``The dog turns around and walks toward the back of the room.''
}
The new action continues from the existing generation state. A second clip begins growing from the end of the first clip on the timeline, and the preview shows the dog turning and walking away. In this way, Eric constructs the shot incrementally, with each Action node specifying what should happen next.

\paragraph{Retaking from an intermediate frame.}
While watching the dog turn, Eric decides that it should stop halfway and turn back toward the camera. He pauses the generation and drags the timeline \textit{playhead} to a frame where the dog has partially completed its turn. He then clicks the \textit{Retake} button.
A new Action node appears on the Shot Canvas at the selected point in the generation history. Eric enters a different continuation:
\textit{
``The dog stops midway, turns back, and looks toward the camera.''
}
The new clip appears on a branch track above the original continuation, beginning at the selected frame. It does not overwrite the footage already generated. Instead, StreamStudio preserves the shared prefix and gradually extends a new continuation from the retake point. Because generation resumes with the inference state associated with the selected frame, the dog's pose and motion remain continuous with the preceding footage.

\paragraph{Exploring another branch.}
Eric then considers another possibility: the dog could leave the scene before turning around. He returns to the first clip, moves the playhead to a frame where the dog is still standing and blinking, and clicks \textit{Retake} again. In the newly created Action node, he enters:
\textit{
``The dog walks toward the right and exits the frame.''
}
StreamStudio generates another clip from this earlier moment. Eric now has two new branches to compare: one in which the dog begins turning away but returns to face the camera, and another in which it walks out of the frame. The original continuation also remains available. These trajectories share the same opening but diverge at different moments, allowing Eric to explore alternative actions without repeatedly regenerating the accepted beginning.
The Shot Canvas displays the relationships among these alternatives, while the timeline places their clips on separate tracks aligned with their corresponding branch points. Eric can switch between the trajectories, compare their timing, and continue generating from any of them.

\paragraph{Editing the appearance of a branch.}
After reviewing the alternatives, Eric decides to make one branch more playful. He selects a frame and clicks the \textit{Edit} button. StreamStudio creates an editing node in the Shot Canvas, where Eric enters:
\textit{
``The dog is wearing a hat.''
}
StreamStudio applies streaming video-to-video editing from the selected frame and produces another branch. The footage before the intervention remains unchanged, while the dog appears with a hat in the subsequent frames. The edited result grows on a separate timeline track, preserving both the source trajectory and the new visual alternative.

\paragraph{Using multiple characters as reference.}
Eric then returns to the Overview Canvas to continue developing the project. He creates two additional Image nodes and generates reference images of a cat and a pig. Next, he creates another Shot node and connects the dog, cat, and pig Image nodes to it as reference images. To retain the room established in the earlier video, he also uses a frame from the dog shot as the first-frame condition.
Eric opens the new Shot and creates an Action node describing the cat and pig entering the room and joining the dog. StreamStudio uses the connected animal images to preserve their appearances and the selected video frame to maintain the existing setting. Eric can then continue generating, retaking, editing, and branching this multi-character scene using the same workflow.

Through this process, Eric moves between two levels of authoring. The Overview Canvas supports project-level composition by organizing reusable assets and shots, while the Shot Editor supports temporal exploration through observable generation, prompt-based continuation, historical retakes, and streaming video editing.

\begin{figure}[h!]
    \centering
    \includegraphics[width=1\linewidth]{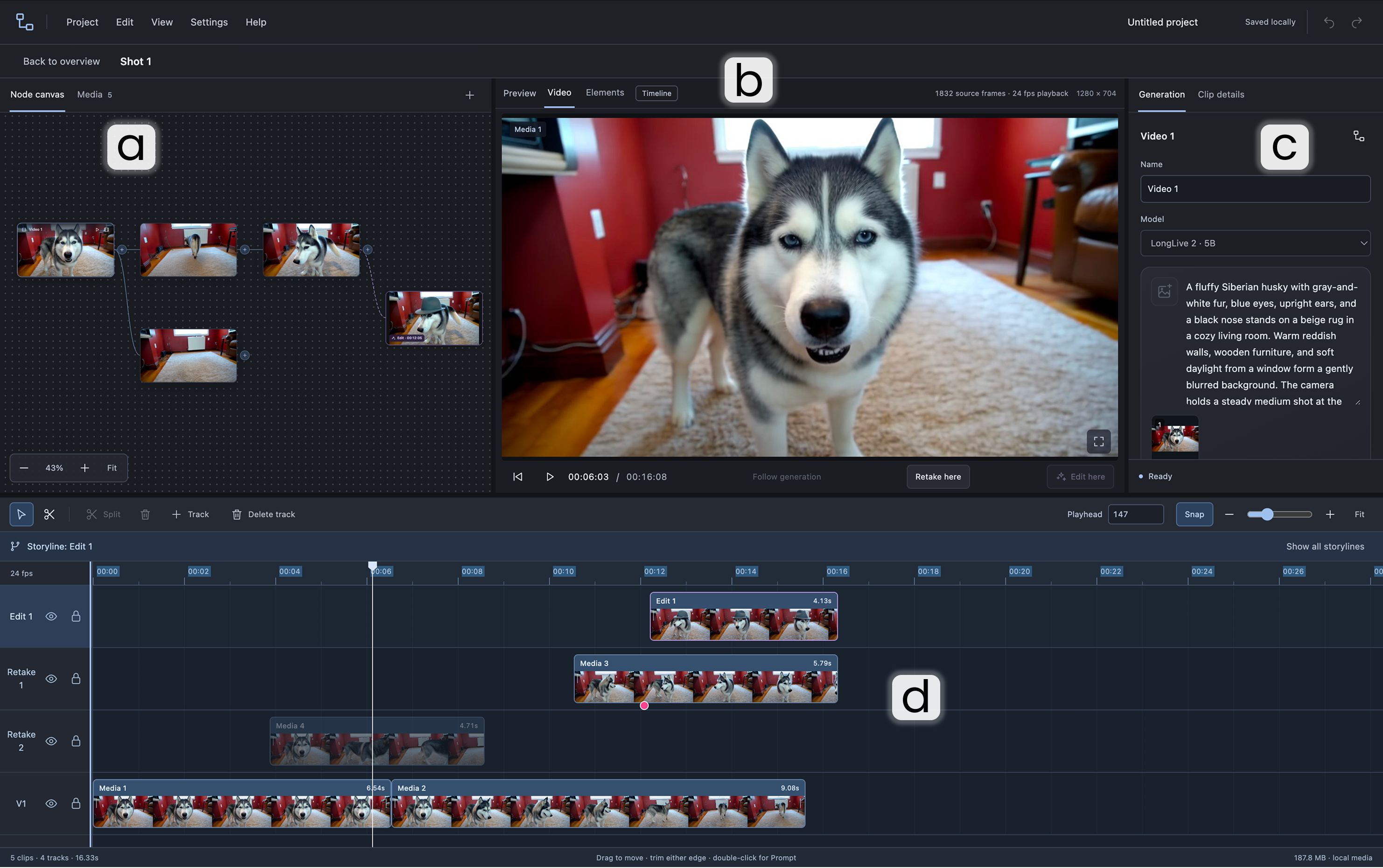}
    \caption{\textbf{StreamStudio's Shot Editor Interface.} (a) The Shot Canvas records Action Nodes and their branching relationships, with each node corresponding to a generated Clip on the Timeline. (b) The Preview displays the frame at the current playhead and follows an ongoing generation. (c) The Action Details panel shows the prompt, model, and generation controls of the selected Action Node. (d) The multi-track Timeline arranges generated Clips and alternative Retake and Edit branches over time.}
    \label{fig:inteface}
\end{figure}

\subsection{StreamStudio Components}
\label{sec:components}

StreamStudio organizes video authoring at two connected levels. The \textit{Overview Canvas} represents the overall project structure, while the \textit{Shot Editor} exposes the temporal generation process within an individual Shot (Fig.~\ref{fig:inteface}). 

\paragraph{Overview Canvas.}
The Overview Canvas is the project-level workspace containing visual assets and Shot Nodes. Users can create or import images, arrange them spatially, and connect them to Shots as first-frame or reference-image conditions. The same image can be reused across multiple Shots, and a Shot can receive multiple image references.

\paragraph{Shot.}
A Shot is the primary video-authoring unit in StreamStudio. Rather than representing a single fixed video, it stores its input conditions, Action Nodes, generated Clips, and alternative Branches. Opening a Shot enters the Shot Editor, while returning to the Overview Canvas collapses these internal alternatives into a single Shot Node.

\paragraph{Shot Editor.}
The Shot Editor integrates three synchronized views: a Preview, an Shot Canvas, and a Timeline. The Preview displays the frame at the current playhead and follows newly generated frames during streaming generation. The Shot Canvas records the logical relationships among Action Nodes, while the Timeline displays their corresponding Clips and Branches over time.

\paragraph{Action Node.}
An Action Node represents a creator instruction at a particular point in the generation process. It contains a text prompt, optional visual conditions, and the temporal state from which generation starts. Connecting a node to the current endpoint continues the active trajectory, while creating one from a historical frame starts an alternative trajectory.

\paragraph{Clip.}
A Clip is the generated media segment associated with an Action Node. While the Action Node specifies what should happen, the Clip contains the resulting frames. During streaming generation, the Clip progressively grows on the Timeline as new frames become available, allowing users to inspect the result before generation is complete.

\paragraph{Branch.}
A Branch is an ordered trajectory through one or more Clips, representing one possible version of a Shot. Different Branches may share an existing prefix and diverge after a Retake or Edit operation. In this representation, Action Nodes capture authoring instructions, Clips capture generated results, Branches capture alternative trajectories, and a Shot contains all alternatives belonging to the same video unit.

\subsection{StreamStudio Authoring Operations}
\label{sec:authoring-operations}

\begin{figure}[h!]
    \centering
    \includegraphics[width=1\linewidth]{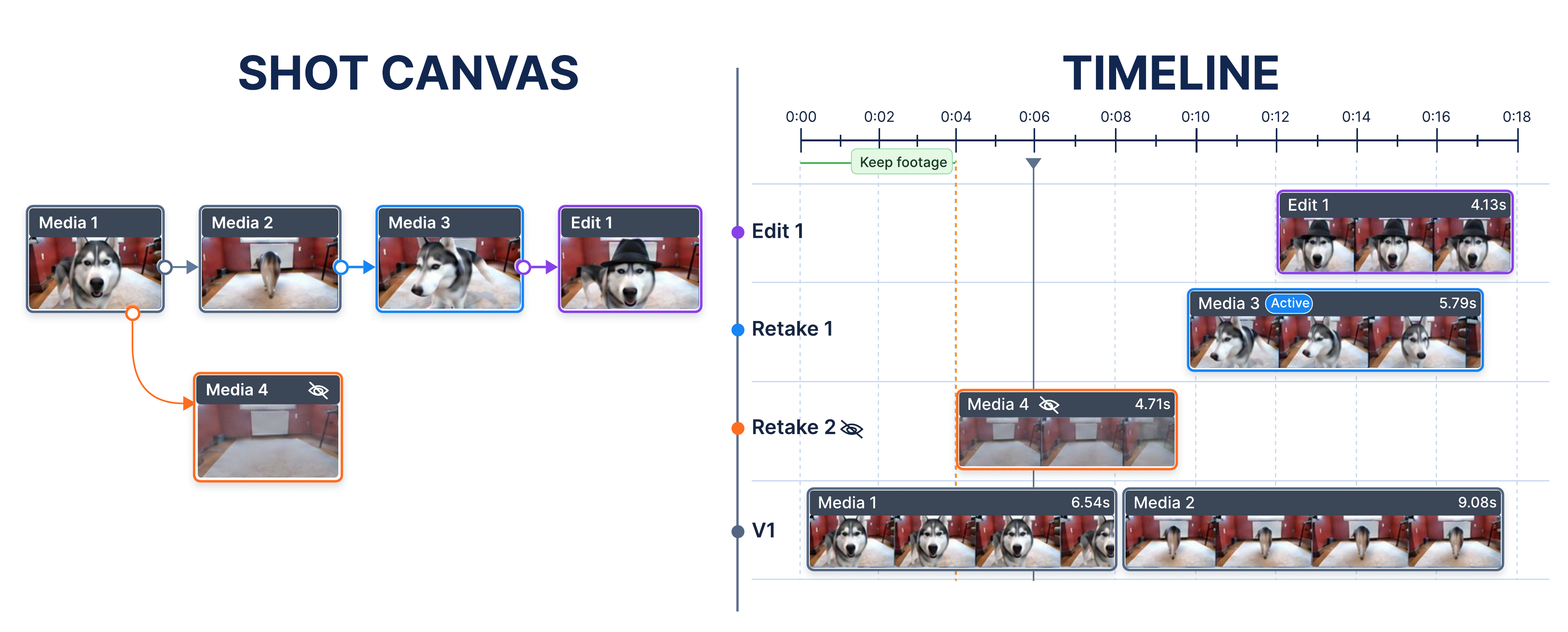}
  \caption{\textbf{Correspondence between the Shot Canvas and Timeline.} Each Action Node on the Shot Canvas corresponds to a generated Clip on the Timeline. Selecting an active Branch displays its associated trajectory, while Clips belonging to other Branches are hidden or visually de-emphasized on their Timeline tracks.}
    \label{fig:shot-canvas-timeline}
\end{figure}

StreamStudio provides four primary operations for constructing and revising a Shot: Generate, Extend, Retake, and Edit. Each operation updates both the Shot Canvas, which records the generation structure, and the Timeline, which displays the resulting Clips (Fig.~\ref{fig:shot-canvas-timeline}).

\paragraph{Generation Conditions.}
An Action Node can use three types of conditioning. A \emph{text prompt} describes the desired scene or action, a \emph{first frame} defines the visual state from which generation starts, and \emph{reference images} provide appearance guidance for characters, objects, or environments without fixing the initial composition. This distinction allows creators to separately control where a generation begins and what visual elements it should preserve.

\paragraph{Generate.}
Generate creates the initial trajectory of a Shot from a text prompt and optional visual conditions. StreamStudio creates a root Action Node and begins synthesizing its Clip, which grows progressively on the Timeline while newly generated frames appear in the Preview. Users can therefore inspect the result and intervene before the full sequence has been generated.

\paragraph{Extend.}
Extend continues generation from the current endpoint of the active Branch. The user creates a new Action Node describing what should happen next, and its Clip is appended to the existing trajectory. This allows a Shot to be constructed incrementally, with each action specified after observing the preceding result.

\paragraph{Retake.}
Retake creates an alternative continuation from a selected historical frame. The user moves the playhead to the desired moment and provides a new instruction, after which StreamStudio creates a new Action Node and Branch from that point. The shared prefix and original continuation are both preserved, allowing alternative actions to be explored without regenerating accepted footage.

\paragraph{Edit.}
Edit modifies the visual content from a selected moment and propagates the change into a new continuation through streaming video-to-video editing. Unlike Retake, which preserves the selected frame and changes what happens afterward, Edit changes the visual state itself and carries that modification forward. The original trajectory remains available for comparison.

Together, these operations allow creators to specify an initial direction, observe generation as it develops, extend the current sequence, or revisit an earlier moment to explore alternative actions and appearances. The following section describes how StreamStudio checkpoints, restores, and forks the underlying causal inference state to support these interactions.

%% file: sections/4_implementation.tex
\section{Implementation}

\subsection{System Setup}
\label{sec:model-setup}

StreamStudio adopts a plug-and-play model interface that separates its authoring operations from specific generative models. The system supports three model-based tasks. Image generation requires a text-to-image model that can optionally condition on reference images and is used to create Image Nodes and first frames. Streaming video generation requires an autoregressive video model that can continue from an initial frame and accept prompt updates during generation. Video editing requires a streaming video-to-video model that follows text instructions while preserving temporal continuity. In our current implementation, we test \texttt{FLUX.2 Klein 4B} for image generation, \texttt{LongLive 2.0 5B} for streaming video generation, and \texttt{JoyAI-Video-Edit} and \texttt{EditStream} for video editing.

\begin{figure}
    \centering
    \includegraphics[width=1\linewidth]{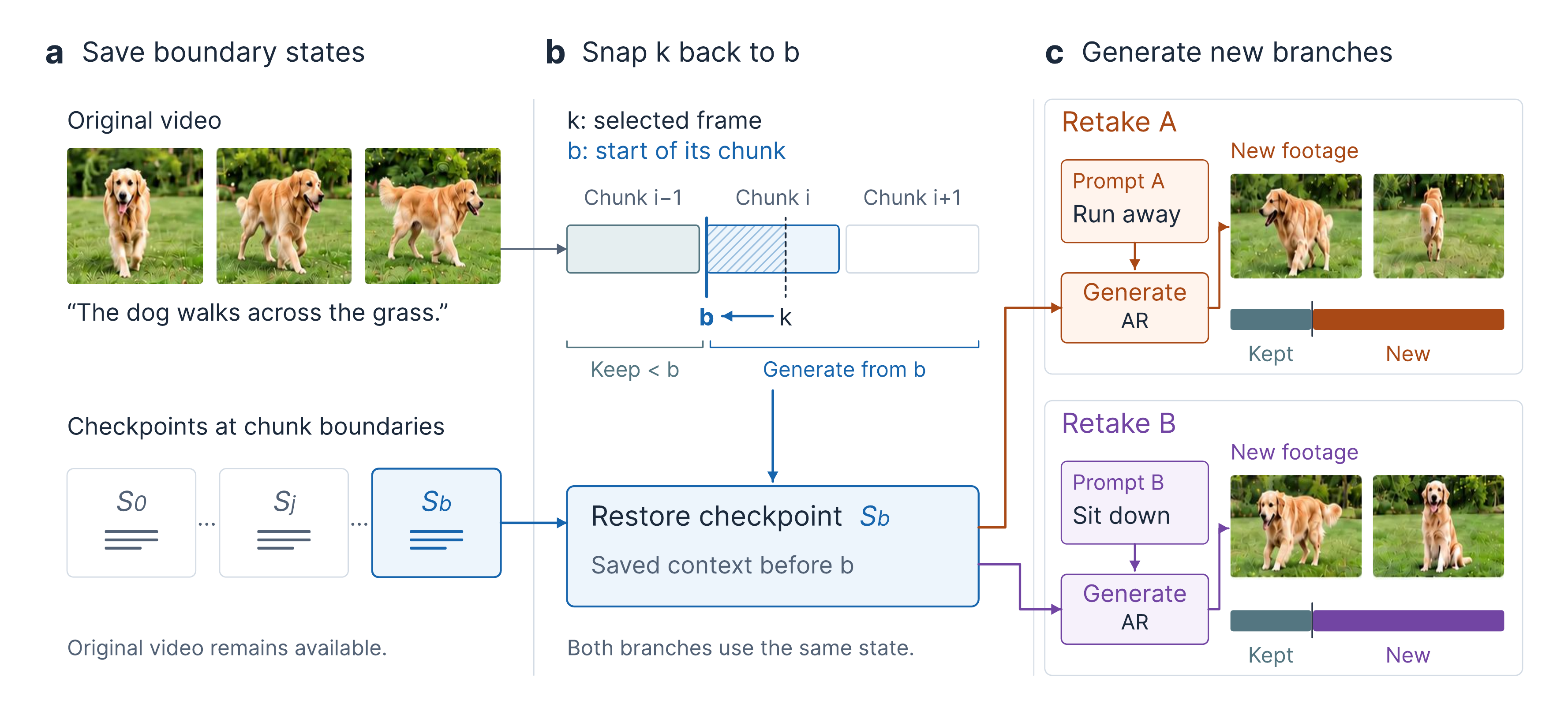}
    \caption{\textbf{Inference state restoration for Retake.} (a) StreamStudio stores an inference-state checkpoint at each chunk boundary while preserving the original video. (b) Given a selected frame $k$, the system maps the branch point to the beginning $b$ of its containing chunk, retains the video prefix before $b$, and restores checkpoint $S_b$. (c) Different prompts fork from the same restored state to generate alternative autoregressive continuations while sharing the retained prefix.}
    \label{fig:retake}
\end{figure}

\subsection{Inference state management}
\label{sec:inference-state-management}

Autoregressive video models generate a video incrementally, with each new chunk conditioned on the prompt and the previously generated chunks. During inference, the model maintains internal states such as the KV cache and latent-frame context, which summarize the preceding visual and temporal information and allow generation to continue without recomputing the entire prefix. We refer to the model-specific information required to resume generation as the \emph{inference state}. This state cannot be recovered from a decoded frame alone, because a frame captures only the visible result rather than the model's internal representation of the preceding motion and context. StreamStudio therefore saves an inference state checkpoint after every generated chunk.

StreamStudio organizes these checkpoints in latent space. Each checkpoint references its parent, while each Branch points to the latest checkpoint in one generation trajectory. Branches with the same prefix reuse the same checkpoint history and store new states only after they diverge. When a user performs a Retake, StreamStudio restores the checkpoint associated with the selected moment, creates a new Branch, applies the new prompt, and continues generation from the restored state (Fig.~\ref{fig:retake}). This design preserves the original continuation, avoids regenerating shared prefixes, and provides the model with the causal context needed to maintain continuity across extensions and Retakes.